\documentstyle[prl,aps,multicol,epsfig]{revtex}              
\begin{document}
\draft
%\begin{frontmatter}
\title{ High-T$_{c}$-Superconductivity and Shadow State Formation \\
        in YBa$_{2}$Cu$_{3}$O$_{6+\delta}$ 
        and Bi$_{2}$Sr$_{2}$CaCu$_{2}$O$_{8+\delta}$}

\author{S. Grabowski, J. Schmalian, and K.H. Bennemann}
\address{Institut f\"ur Theoretische Physik,
  Freie Universit\"at Berlin, Arnimallee 14, \\
       14195 Berlin, Germany}
\date{August 15, 1995}
\maketitle 
\begin{abstract} 
\leftskip 54.8pt
\rightskip 54.8pt
The normal and superconducting state of YBa$_{2}$Cu$_{3}$O$_{6+\delta}$ and 
Bi$_{2}$Sr$_{2}$CaCu$_{2}$O$_{8+\delta}$ are investigated by using the 
mono- and bilayer Hubbard model within the fluctuation exchange 
approximation and a proper description of the Fermi surface topology. 
The inter- and intra-layer interactions, the renormalization of the bilayer 
splitting and the formation of shadow bands are investigated in detail. 
Although the shadow states are not visible in the monolayer, we find that the 
additional correlations in bilayers boost the shadow state intensity and will 
lead to their observability. In the superconducting state we find a $d_{x^2-y^2}$ 
symmetry of the order parameter and demonstrate the importance of inter-plane 
Copper pairing.
\end{abstract}   
\begin{multicols}{2}   
\newpage

Central questions in the theory of the high-$T_c$ superconductors are related to the  
importance of the multiple $CuO_{2}$ layers within the High-T$_c$ superconductors like 
Bi$_{2}$Sr$_{2}$CaCu$_{2}$O$_{8+\delta}$ (BSCCO) or YBa$_{2}$Cu$_{3}$O$_{6+\delta}$ (YBCO)
and its influence on the superconducting pairing symmetry. In addition, the
detailed shape of the Fermi surface (FS) is believed to be of importance for a quantitative
description of transport and photoemission experiments and also for the material dependence
of the transition temperature $T_c$~\cite{levinold,pines}. For bilayer cuprates, neutron scattering 
experiments found indications that there is an antiferromagnetic coupling between 
nearest-neighbor layers in YBCO~\cite{tranquada,fong}. Furthermore, recent angular resolved 
photoemission (ARPES) experiments found evidence for two separated bands in YBCO~\cite{liu,GCA94}, 
that might be related to the existence of two $CuO_{2}$ bands caused by a inter-plane quasi 
particle transfer. However, the small experimentally observed bilayer splitting in YBCO and the 
difficulty to resolve two $CuO_{2}$ bands in BSCCO~\cite{ding,shen} support the idea that 
the strong short ranged antiferromagnetic order in the cuprates reduces the inter-layer 
hopping and might be responsible for the observation of shadows of the FS~\cite{aebi,RKK96} in BSCCO. 

In this paper we study the bilayer Hubbard Hamiltonian within the fluctuation exchange
(FLEX) approximation~\cite{bickers} and focus our attention in particular
on YBCO and BSCCO compounds. We find that the antiferromagnetically correlated planes yield 
strong deformations of the quasi particle dispersions and suppression of the bilayer 
splitting. In addition we observe that the shadows of the FS occur only when the inter-plane 
coupling is considered. The superconducting state is investigated within the framework of the
Eliashberg theory where the order parameter is found to have a $d_{x^2-y^2}$ symmetry with 
inter- and intra-layer Cooper pair formation. 
 
Our theory is based on the general multilayer Hubbard model that will be later on 
specified for a bilayer system:  
\[ 
H=\sum_{i,j,l,l',\sigma} (t^{i,l}_{j,l'}-\mu \; \delta^{i,l}_{j,l'}) \; 
c^{\dagger}_{i,l,\sigma} c_{j,l',\sigma}
+U \sum_{i,l} n_{i,l,\uparrow} n_{i,l,\downarrow}\;,
\]
where the hopping integrals $t^{i,l}_{j,l'}$ determine the bare dispersion 
$\varepsilon^{ll'}_{{\bf k}}$ in 2D ${\bf k}$-space, $i$ and $j$ ($l$ and $l'$) are the site 
(layer) indices, $\delta^{i,l}_{j,l'}$ the Kronecker symbol, $U$ the local Coulomb repulsion and 
$\mu$ the chemical potential. For the bilayer the interaction-free contribution of the Hamiltonian 
can be diagonalized with the unitary transformation ${\cal U}$
yielding an antibonding ($-$) and a bonding band ($+$) with bare dispersion 
$\varepsilon_{{\bf k}}^{\pm}$. Assuming that this symmetry holds also in the full interacting 
case, one can define for the normal state corresponding Greens function $G_{\pm} ({\bf k},i\omega_m)$ 
with fermionic Matsubara frequencies $\omega_m=(2m + 1)\pi T$ and temperature $T$. 

In the superconducting state it is helpful to investigate the allowed symmetries of the 
order parameter in a multilayer system. Here we consider only singlet superconductivity and 
for simplicity the bilayers case although similar symmetries apply to the multilayer Hamiltonian. 
Note that this discussion is only related to the inter-layer effects and not restricted to a
certain in plane symmetry ($d$- or $s$-wave). At first by interchanging the two electrons of the 
Cooper pair, it follows for the gap function in the layer representation that 
$\Delta_{l l'}({\bf k})= \Delta_{l' l}(-{\bf k})$ due to the Pauli principle. Note that the 
frequency indices have been omitted for clarity. In addition to the in-plane symmetries, 
present for a single layer, the Hamiltonian is invariant with respect to the inversion 
symmetry with inversion center between the layers: therefore, the gap function has a given 
parity $P=\pm1$ and it follows $\Delta_{11}({\bf k})=P \Delta_{22}(-{\bf k})$ and
$\Delta_{12}({\bf k})=P \Delta_{21}(-{\bf k})$.
Using the in-plane symmetry ${\bf k} \rightarrow -{\bf k} $ it follows from this considerations
that the gap-function for even parity  pairing $P=+1$ is given in the layer 
representation by:
\begin{eqnarray}
\Delta({\bf k})  =   \left(\begin{array}{cc}
\Delta_{||} ({\bf k})  &   \Delta_{\perp} ({\bf k}) \\
\Delta_{\perp} ({\bf k})           &   \Delta_{||} ({\bf k})
\end{array} \right)  \, .
\label{even}
\end{eqnarray}
Here, the gap-function can be diagonalized by the transformation ${\cal U}$ of the normal 
state and  intra-band pairing  (simultaneous intra- and interlayer pairing) occurs.
For odd parity pairing $P=-1$ it follows similarly for the gap function
\begin{eqnarray}
\Delta({\bf k})  =   \left(\begin{array}{cc}
\Delta_{||} ({\bf k})  &   0 \\
           0           &  - \Delta_{||} ({\bf k})
\end{array} \right)  \, ,
\label{odd}
\end{eqnarray}
and only intra-layer pairing occurs. Hence the gap function is off diagonal in the eigenvalue 
representation leading to an interband pairing state. Consequently, the corresponding Eliashberg
equations decouple and the symmetry with the larger $T_c$ will determine the superconducting state.
However one might argue that there will be a change from intra- to inter-band pairing or vise
versa for $T<T_c$ which would be related to a second phase transition below $T_c$. So far this
has not been observed experimentally, and we will only consider the solution with the larger
energy gain of the condensate. Now, by looking at the inter-band pairing, for a given momenta
two electrons from different bands that form a Cooper pair cannot origin both from the FS
due the structure of the bonding and antibonding band and due to the bilayer splitting. 
Hence, we expect intra-band pairing to be most dominant, e.g. yielding the largest energy 
gain of the superconducting phase which will be treated in the following. Note that this point
is also supported by weak coupling approaches as shown by Maly {\em et al.}~\cite{intra}.
 
Now to treat the superconducting state of the multiband Hubbard Hamiltonian we use the 
Nambu-Eliashberg approach where the Green's function and the off-diagonal Green's function that 
describes superconducting correlations are written in terms of $2 \times 2$ matrices. Here we have 
to account for an additional index for the layer or the eigenvalue representation~\cite{mont,supra}. 
Consequently the Eliashberg equations for intra-band pairing can be written as follows by expanding 
the matrix self energy in terms of Pauli matrices:
\begin{eqnarray}
 \Phi_{{\bf k}, \lambda}(i\omega_n) &=&\frac{T}{N} \sum_{{\bf k}', \lambda',n'}
\frac{\left({\tilde V}^{\lambda,\lambda'}_{{\bf k}-{\bf k}'}(i\omega_n-i\omega_{n'})  
+U \delta_{\lambda,\lambda'} \right)} {D_{{\bf k}' , \lambda'}(i\omega_{n'})} \nonumber  \\
& &\times \Phi_{{\bf k}', \lambda'}(i\omega_{n'}) \, , \nonumber   \\
%\end{eqnarray}
%\begin{eqnarray}
  X_{{\bf k}, \lambda}(i\omega_n) & = &\frac{T}{N} \sum_{{\bf k}', \lambda',n'}
\frac{V^{\lambda,\lambda'}_{{\bf k}-{\bf k}'}(i\omega_n-i\omega_{n'})}  
 {D_{{\bf k}' , \lambda'}(i\omega_{n'})} \nonumber  \\
& &  \times (\varepsilon_{{\bf k}' \lambda'} +X_{{\bf k}', \lambda'}(i\omega_{n'}))    \, , 
\nonumber  
\end{eqnarray}
\begin{eqnarray}
i\omega_n(1-Z_{{\bf k}, \lambda}(i\omega_n)) &=& \frac{T}{N} \sum_{{\bf k}', \lambda',n'}
\frac{V^{\lambda,\lambda'}_{{\bf k}-{\bf k}'}(i\omega_n-i\omega_{n'}) } {D_{{\bf k}' , \lambda'}(i\omega_{n'})}
\nonumber \\
& & \times i\omega_{n'}Z_{{\bf k}', \lambda'}(i\omega_{n'})  \, , \nonumber  
\end{eqnarray}
where 
\begin{eqnarray}
D_{{\bf k}' , \lambda'}(i\omega_{n'})& = &
(i\omega_{n'} Z_{{\bf k}', \lambda'}(i\omega_{n'}))^2 \nonumber \\
& &-(\varepsilon_{{\bf k}',\lambda'}
-X_{{\bf k}', \lambda'}(i\omega_{n'}))^2 -
\Phi_{{\bf k}', \lambda'}(i\omega_{n'})^2 .\nonumber
\end{eqnarray}
Here the term $U \delta_{\lambda,\lambda'}$ accounts for the Hartree contribution which is for the 
diagonal elements absorbed in the chemical potential and $\varepsilon^{\lambda}_{{\bf k}}$ is the 
free dispersion that determines the FS shape. The expansion coefficients of the diagonal self energy 
are $i\omega_{n} (1-Z_{{\bf k},\lambda}(i\omega_{n}))$ and $\chi_{{\bf k},\lambda}(i\omega_{n})$, 
whereas $\phi_{{\bf k},\lambda}(i\omega_{n})  = \Delta_{{\bf k},\lambda}(i\omega_{n})   
Z_{{\bf k},\lambda}(i\omega_{n}) $ is the coefficient of the off-diagonal self energy which signals 
superconductivity and $\Delta_{{\bf k},\lambda}(i\omega_{n})$ is the gap function. The interactions 
$V^{\lambda,\lambda'}_{{\bf q}}$ and ${\tilde V}^{\lambda,\lambda'}_{{\bf q}}$ for the bilayer systems 
can be obtained by performing the summation of the FLEX diagrams~\cite{bickers} in the layer representation 
and are similar to the monolayer case when one considers the fact that a third dimension with two
momenta points $\pi$ and $0$ is introduced. Furthermore for the bilayer system they have an 
inter-plane, $V^{\bot}_{{\bf k}} (i\omega_m)$, and an in-plane, 
$V^{\parallel}_{{\bf k}}(i\omega_m)$, contribution:
\begin{eqnarray}
V^{++}({\bf k},i\omega_m)&=&
1/2\;(V^{\parallel}_{{\bf k}} (i\omega_{n}) \;+ V^{\bot}_{{\bf k}} (i\omega_{n})) 
\nonumber\\ 
V^{+-}({\bf k},i\omega_m)&=&
1/2\;(V^{\parallel}_{{\bf k}} (i\omega_{n}) \;- V^{\bot}_{{\bf k}} (i\omega_{n})) \;.
\label{trafo} 
\end{eqnarray}

This set of coupled Eliashberg equations is solved self-consistently on the real frequency 
axis~\cite{details}. Since most photoemission experiments were performed on YBCO and in particular 
BSCCO systems, it is necessary for a detailed comparison of our theory with experiments to use 
dispersions and Fermi surfaces that are closed around the ($\pi,\pi$) point. Thus the FS topology is 
characterized by 
\begin{eqnarray}
\varepsilon^{\pm}_{{\bf k}}&=&-[2t(\cos (k_x)+\cos (k_y)) + 4t'\cos (k_x) \cos (k_y)
\nonumber\\
&& +2t'' (\cos (2k_x)+\cos (2k_y))\pm t_{\bot}] 
\nonumber
\end{eqnarray}
with intra-plane hopping integrals $t=0.25$ eV, $t'=-0.38t$, $t''=-0.06t$ and 
$t_{\bot}=0.4t = 100$~meV~\cite{ALJ95} as explicit model for YBCO.  Due to the similarity of 
the YBCO and BSCCO FS, we used for the later one the bare intra-plane dispersion of YBCO and 
$t_{\bot}({\bf k})=1/4 \; t_{\bot}\;(\cos (k_x)-\cos (k_y))^{2}$ with $t_{\bot}=0.4t$. The 
resulting FS is very similar to the experiments~\cite{shen} and to the band structure
calculations~\cite{massida}. For comparison with previous results we take $U=4t$ but notice that 
we find no significant changes in our data up to values of $U = 6t$.  
 
In Fig. 1 we present out data for the effective interaction $V_{{\bf q}}(\omega)$ in the two
dimensional Brillouin zone (BZ) for $\omega=0$ where we have for simplicity neglected the bilayer 
coupling ($t_{\bot}=0$). Note that $x=1-n$ is the doping concentration whereby $n$ is the occupation 
number per site. Interestingly, we find commensurate peaks in the effective interaction for this compound 
as observed in neutron scattering experiments~\cite{tranquada}. This results is caused by the electronic 
correlations since for $U=0$ pronounced incommensurabilities are present due to the fact that 
$2\;{\bf k_{FS}} \neq (\pi,\pi)$ with Fermi surface momentum ${\bf k_{FS}}$ that shift the peaks 
in $V_{{\bf q}}(\omega)$ away from ${\bf Q}$. However for finite $U$ and strong scattering rates these 
incommensurable structures are smeared out due to the overdamped nature of the spin excitations and a 
peak at ${\bf Q}$ remains. So far only the in-plane correlations in an insulated $CuO_2$-plane are 
considered. However, when various layers are coupled like in the bilayer systems YBCO and BSCCO the 
influence of the inter-plane interaction $V^{\bot}_{{\bf q}}(\omega) $ on the superconducting state and 
$T_c$ is a significant and important open question. By investigating $V^{\bot}_{{\bf q}}(\omega)$ for the 
YBCO and BSCCO parameterization with $t_{\bot}=0.4t$ we find  that the commensurabillities in 
$V^{\parallel}_{{\bf q}}(\omega) $ remain or get even more pronounced while $V^{\bot}_{{\bf q}}(\omega)$ 
has the same ${\bf k}$-dependence as $V^{\parallel}_{{\bf q}}(\omega)$ but with an opposite sign. This 
yields an antiferromagnetic correlated bilayer via the hopping $t_{\bot}$ that is strongest for low doping 
concentrations, where inter- and intra-layer interactions become comparable in magnitude although we 
always find $V^{\bot}_{{\bf q}}(\omega) < V^{\parallel}_{{\bf q}}(\omega)$.

The formation of shadow states is an important issue in the recent research, since it concerns the 
transition of the quasi particle excitations with doping from the simple paramagnetic state for large
$x$ to the antiferromagnet for low $x$ and from large FS for optimally doped systems to small FS hole
pockets for low doping. Recently we discussed for a simple model dispersion with $t'=t''=0$ that the 
dynamical antiferromagnetic short range order in the cuprates leads to a transfer of spectral weight 
from the FS at ${\bf k_{FS}}$ to its shadow at ${\bf k_{FS}+Q}$~\cite{shadow} that might lead to the observed ARPES spectra
~\cite{aebi,RKK96}. In this context the importance of the quasi two dimensional nature of the magnetic
excitations for shadow states has been discussed recently in Ref.~\onlinecite{vilk1}. Now, in Fig. 2 we 
demonstrate the influence of the realistic YBCO or BSCCO dispersion on the shadow states for $t_{\bot}=0$. 
Here we plot the spectral density $\rho_{{\bf k}} (\omega)$ for two doping concentrations and for
${\bf k}$ near the crossing of the dispersion with the shadow of the FS near ($\pi/2,\pi/2$) (a), for 
${\bf k}$ at the FS near ($\pi,0$) (b) and for ${\bf k}$ at the shadow of the FS near ($\pi,0$) (c-d).
Qualitatively, these data agree with our findings for the previously used model compound since we
find pronounced occupied and unoccupied shadow states in the BZ as small additional satellites in the
spectral density. Again, these states are not related to new quasi particles but rather to an
incoherent amount of spectral weight caused by an increased scattering rate~\cite{shadow}.
However by comparing the data for $x=0.02$ and $x=0.08$ as in Fig. 2~(c) and (d), we see 
that they become much weaker due less pronounced nesting of the FS and the corresponding much smaller 
magnetic interaction $V_{{\bf q}}(\omega)$ as can be seen in the inset and in Fig. 1. Comparing these
results with the $t'=t''=0$ FS we find that for $x=0.08$ the shadows are rather strong~\cite{shadow}.
Thus the observation by Aebi {\it et al.}~\cite{aebi} can not be satisfactorily understood by considering 
only a single $CuO_{2}$ plane which is in agreement with the study of Ref.~\onlinecite{vilk2}.
since the shadows were found near the optimal doping at $x \approx 0.15$.
 
However, as recently discussed for the $t'=t''=0$ model dispersion~\cite{bilayer} the consideration
of a finite $t_{\bot}$ increases the shadow state intensity due to the additional bilayer correlations.
Here the fact that for low doping $V^{\bot}_{{\bf q}}(\omega) \approx V^{\parallel}_{{\bf q}}(\omega)$ 
lead via Eq.~\ref{trafo} to $V^{++}_{{\bf q}}(\omega) \approx 0$ and  
$V^{+-}_{{\bf q}}(\omega) \approx V^{\parallel}_{{\bf q}}(\omega)$ such that we find that the spectral 
weight is not only shifted by the momentum ${\bf Q}$, but simultaneously also from the bonding to the 
antibonding band and vise versa. Now by taking a finite $t_{\bot}$ into account we find for YBCO and 
BSCCO-like systems that the inter-layer antiferromagnetic coupling also increases the shadow state 
intensity and they start to appear for $x<0.12$ with a maximum intensity at $t_{\bot} \approx 0.4t$. 
Furthermore as demonstrated in Fig. 3 for YBCO we find that the most favorable region to observe shadow 
states in ARPES is in the neighborhood of the ($\pi/2,\pi/2$) point, where main and shadow band are well 
separated. Near ($\pi,0$) the absolute intensity of the shadow states is largest, but they are difficult 
to detect because of the superposition of shadow peaks and the dominant main band contributions. Furthermore 
the strongest shadow band signal might not be obtained with photoemission but with an inverse photoemission 
measurements (IPES) as shown in Fig. 3 (a) where a pronounced shadow peak appears. However the current
limited resolution of IPES measurements might impede the observability of these states. For the BSCCO 
system where we use a ${\bf k}$-dependent inter-layer hopping $t_{\bot}({\bf k})$ as suggested be
band structure calculations~\cite{massida}, we find similar results near ($\pi,0$). However on the 
diagonal at ($\pi/2,\pi/2$) both bands are not splitted and the shadows of the bonding and antibonding 
band are at the same position.

The overall shape of the quasi particle dispersion in the bilayer compound YBCO and the shadow band 
formation is presented in Fig. 4 where we focus our attention on the bilayer splitting.  
The experimental findings concerning the bilayer splitting in the cuprates are still controversial. For 
YBCO Liu {\em et al.}~\cite{liu} and Gofron {\em et al.}~\cite{GCA94} found indications via ARPES for a 
splitting $\Delta^{exp} \varepsilon (\pi,0) = 110$~meV that is much smaller that the theoretical predictions 
from band structure calculations~\cite{ALJ95} whereby for BSCCO different photoemission groups come to 
different conclusions concerning the existence or the absence of a finite bilayer splitting~\cite{ding,shen}.
However, from our results for the bilayer Hubbard model we conclude that the strong antiferromagnetic 
correlations in the high-T$_c$ superconductors are responsible for the smallness of the bilayer splitting 
and might explain these interesting experimental results as also discussed by Liechtenstein {\em et al.}
~\cite{licht}.
 
At this stage, it is of interest to compare our results with the interesting argumentation of Vilk
~\cite{vilk2} that shadow states and a variety of related phenomena appear only if the magnetic
correlation length $\xi$ is larger than the thermal de Broglie wavelength $\lambda_{th}=v_F/\pi k_{B} T$.
Here $v_F$ is the Fermi velocity of the system. Assuming that $v_F$ is given by its uncorrelated value
$v_F \approx \pi a \;t$ ($a$: lattice constant) would yield shadow states for low temperatures only for 
extremely large magnetic correlation length. However, our self consistent calculations yields a self 
stabilization of the shadow band phenomena: the increasing quasi particle scattering rate which is related 
to the formation of shadow states~\cite{shadow} gives rise to a pronounced flattening of the correlated 
dispersion as can be seen in Fig. 4. This causes a renormalization and a decrease of the Fermi velocity
for decreasing $T$. This mechanism allows observable shadow bands even for low temperatures since 
$\lambda_{th}$ remains almost constant as a function of $T$ and consequently $\lambda_{th} < \xi$ can be 
fulfilled for moderate correlation length as observed experimentally in the high-$T_c$ superconductors. 
Note that the limit $T \to 0$ is still an open question although we believe that for very low temperatures 
the susceptibility and the correlation length $\xi$ behave differently compared to the effective 
interaction in the expression for the self energy~\cite{diez}. Thus this demonstrates impressively the 
importance of a self consistent calculation of the Fermi velocity that decreases down to $T \approx 40$~K 
by taking explicitly the changes of the spin fluctuation and the quasi particle spectrum into account.
 
In Fig. 5 we present our data for the superconducting order parameter for the YBCO model where we plot
the frequency dependence of the superconducting order parameter $\phi_{{\bf k}} (\omega)$ at 
${\bf k}=(\pi,0)$ for $x=0.08$. Here we treated the mono and bilayer Hubbard model below $T_c$ by solving 
the strong coupling Eliashberg equations thereby considering the full momentum, frequency and doping 
dependence of the intra- and inter-layer interactions. Here one obtains one order parameter for each band, 
namely $\phi^{\pm}_{{\bf k}} (\omega)$. These are connected to the layer representation via 
$\phi^{\pm}_{{\bf k}} (\omega)=\phi^{\parallel}_{{\bf k}} (\omega) \pm \phi^{\bot}_{{\bf k}} (\omega)$, 
where $\phi^{\parallel}_{{\bf k}} (\omega)$ ($\phi^{\bot}_{{\bf k}} (\omega))$ describes intra- (inter-) 
layer Cooper pair formation. By solving these equations for the YBCO dispersion with $t_{\bot}=0$, we 
find for all $x$ a $d_{x^2-y^2}$ wave superconducting state ($T_c=70$ K for $x=0.08$) without the need 
to introduce phenomenological interactions that enforce the formation of Cooper pairs~\cite{dahm1,dahm2}. 
Note that this value is only slightly smaller than the corresponding value for $t'=t''=0$ ($T_c=97$~K for 
$x=0.12$) whose FS is rather similar to the La$_{2-x}$Sr$_{x}$CuO$_{4}$ (LSCO) compound. Concerning the 
pairing symmetry in a bilayer system also find $d_{x^2-y^2}$ pairing symmetry and similar values for the 
transition temperatures. Interestingly, as can be observed in Fig. 5 the $d$-wave state is characterized 
by an increasing contribution of coherent inter-layer pairing with decreasing doping, where for $x=0.08$
Cooper pairs are formed by electrons from the same layer as from different layers with almost equal 
probability. 

Although it is a remarkable success that spin fluctuation induced pairing interaction cause
superconductivity in mono and bilayer compounds, it might be even more important to explain the
rather different transition temperatures in the large family of the high-$T_c$ superconductors. Here
a comparison of our monolayer results for LSCO with the data of the bilayer YBCO compound shows  
that $T_c$ of YBCO should be even slightly lower than for LSCO the system when all other parameters 
like $U$ or the hopping $t$ are taken to be the same which is definitely in contradiction to the 
experimental observations ($T_c^{YBCO}\approx 90$~K versus $T_c^{LSCO} \approx30$~K). 

However it has been recently argued that the larger $T_c$ is YBCO is mainly due to finite size
effect and to the relatively strong coupling of the unit cells~\cite{SGC91}, which is not included 
in our approach.  This point is also supported by experiments on ultrathin YBCO layers separated by 
insulating PrBa$_{2}$Cu$_{3}$O$_{7-\delta}$ which show that one separated bilayer of YBCO yields 
$T_c\approx 20$~K and that the coupling of these bilayers boost the transition temperature to its bulk 
value~\cite{moybco1,moybco2}. In this context our findings that a single bilayer of YBCO has a smaller 
$T_c$ that LSCO are in qualitative agreement with the experiment although there is still the need for 
a consideration of inter-bilayer effects and for understanding the importance of other ingredients of 
the high-$T_c$ materials for superconductivity
 
In conclusion, we investigated  the doping dependence of the mono- and bilayer Hubbard for the YBCO and
BSCCO systems by using the FLEX approximation. The strong antiferromagnetic coupling between the layers 
leads to an enhanced shadow state intensity and to their observability in YBCO- and BSCCO-like models. We 
found that the bilayer splitting is reduced ($\sim 50 \% $) and the inter-layer hopping is effectively
blocked for small doping. Finally, we presented results for the superconducting which has a $d$-wave 
order parameter for mono- and bilayer compounds whereby the later one has significant inter-layer Cooper pair
contributions.

%
%
%----References
%
%

 \end{multicols}
%\newpage
\begin{figure}
\centerline{\epsfig{file=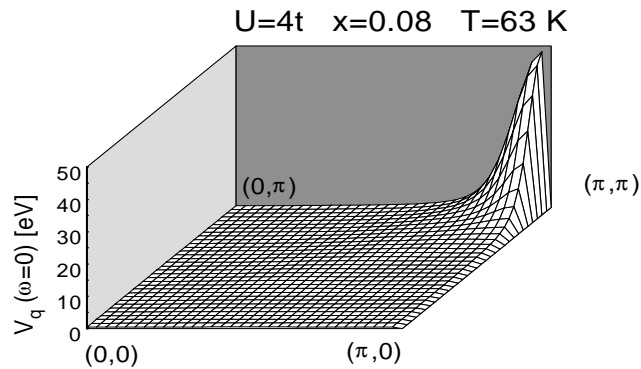,width=9cm,height=11cm}}
\caption{Effective interaction $V_{{\bf q}} (\omega=0)$ of the BSCCO/YBCO model. Note 
the commensurable structures, e.g. the fact that $V_{{\bf q}}(\omega=0)$ is peaked 
at ($\pi,\pi$) as observed in experiments. }
\end{figure}

\begin{figure}
\centerline{\epsfig{file=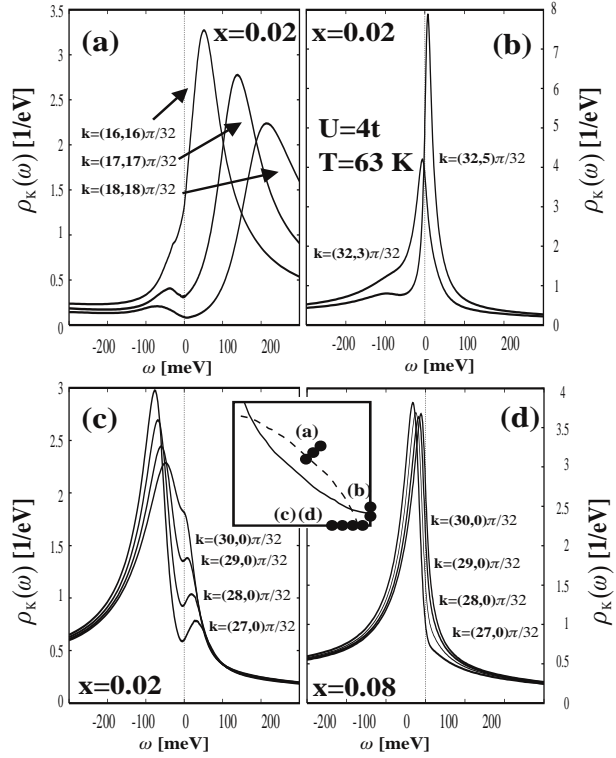,width=8cm,height=10cm}}
\vskip 1cm
\caption{Spectral density $\rho_{{\bf k}} (\omega)$ for the BSCCO/YBCO dispersion for 
${\bf k}$ points as indicated in the inset and two doping concentrations. (a): Crossing 
of the shadow of the Fermi surface on the diagonal in the BZ. (b): Fermi surface crossing 
near ($\pi,0$). (c) and (d): Crossing of the shadow of the Fermi surface near ($\pi,0$)
for $x=0.02$ and $x=0.08$.} 
\end{figure}

\begin{figure}
\centerline{\epsfig{file=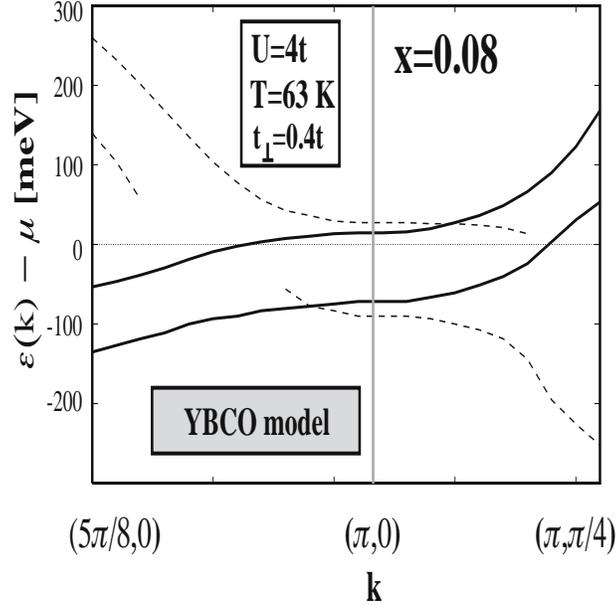,width=8cm,height=8cm}}
\vskip 1cm
\caption{Formation of shadow states for YBCO. (a): $\rho^{-}_{{\bf k}} (\omega)$ near 
($\pi,0$). (b): $\rho^{-}_{{\bf k}} (\omega)$ on the diagonal near ($\pi/2,\pi/2$).
For simplicity only the FS of the bonding band and the shadow of the antibonding 
band are shown in the inset.} 
\end{figure}

\begin{figure}
\centerline{\epsfig{file=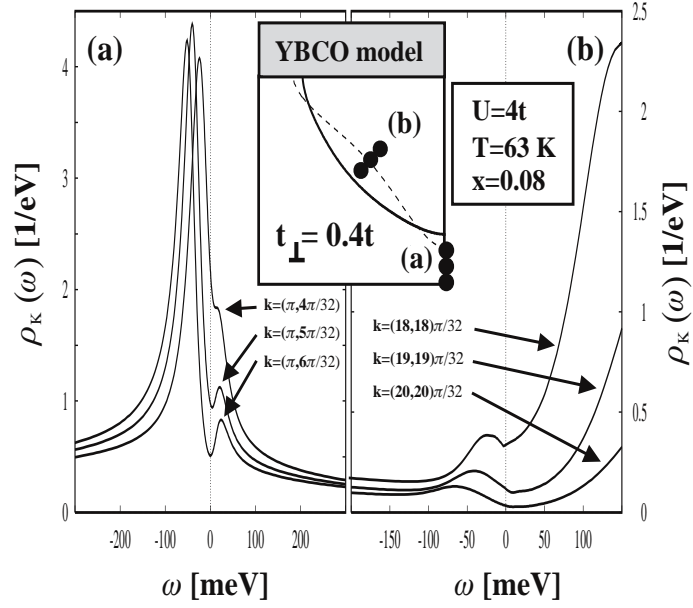,width=9cm,height=8cm}}
\vskip 1cm
\caption{Quasi particle dispersion of the bilayer BSCCO and the YBCO model (solid lines: main bands,
dashed lines: shadow bands). Note the strong suppression of the bilayer splitting from 
$\Delta \varepsilon = 200$~meV for $U=0$ to $\Delta \varepsilon_{(\pi,0)} = 86$~meV.}
\end{figure}
 
\begin{figure}
\centerline{\epsfig{file=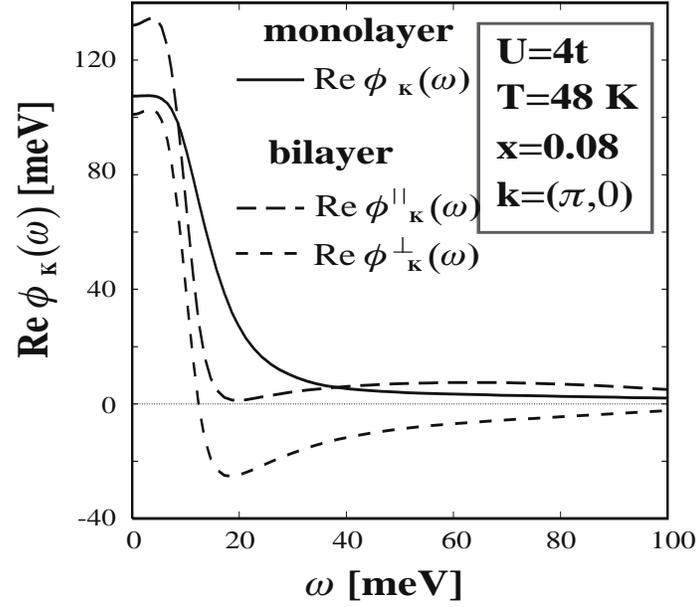,width=9cm,height=8cm}}
\vskip 1cm
\caption{Superconducting order parameter ${\rm Re}\;\phi_{{\bf k}} (\omega)$ for the monolayer
($t_{\bot}=0$) and for the bilayer with $t_{\bot}=0.4\;t$ at ${\bf k}=(\pi,0)$ and the YBCO 
dispersion. Note the large contribution of the inter-layer Copper pair formation.} 
\end{figure}

\end{document}